# Realization of A Strong Ferroelectric Metal by Nb-doping in Strained EuTiO$_3$

Sheng Xu,[1] Yanni Gu,[1] Chuanbo Zheng,[1] Xiao Shen[2]

[1]School of Metallurgy and Materials Engineering, Jiangsu University of Science and Technology, Zhangjiagang, Jiangsu 215600, China

[2]Department of Physics and Materials Science, University of Memphis, Memphis, Tennessee 38152, USA

**ABSTRACT**

Ferroelectric (FE) metals have been attracting attention as they possess both metallicity and ferroelectricity, the two seemingly incompatible physical properties. An important problem for both fundamental research and potential applications is how to realize a strong FE metal. One strategy is to dope a strong FE insulator into an FE metal provided that the induced free electrons do not destroy FE distortion. Strained EuTiO$_3$ film is a strong FE ferromagnet and a promising candidate for such a strategy. Here, we calculate the structural, electronic, and polarization properties of Nb-doped strained EuTiO$_3$ film using the hybrid density-functional theory. The results show that the strained EuTi$_{0.875}$Nb$_{0.125}$O$_3$ film is a strong FE metal as the Nb doping induces metallicity without weakening the FE distortion. The underlying atomistic mechanism of the coexistence of metallicity and strong ferroelectricity is discussed. The findings show that combining doping and strain engineering is a promising way to realize new EuTiO$_3$-based strong FE metals and may be used in other materials as well.

Ferroelectricity cannot exist in ordinary metals because the conducting free electrons can screen static internal electric fields. However, half a century ago, Anderson and Blount[1] proposed so-called polar or ferroelectric (FE) metal in the case of martensitic transformations, provided that free electrons are very weakly coupled to the FE distortions. The first experimental confirmation of their proposal was found in LiOsO$_3$,[2] where the material exhibits a ferroelectric-like structural phase transition at ∼140 K. Since then, FE metals have attracted great attention, and a number of FE metal materials have been discovered via experimental and theoretical methods.[3–7] These polar or FE metals have potential applications in FE devices,[8] nonlinear optics,[9] and as topological materials for quantum devices.[10]

An important problem for fundamental research and potential applications is how to realize a strong FE metal, aka an FE metal with large polar distortions. To achieve this goal, it is important to examine the strategies to realize new FE metals in general. Currently, three strategies have been adopted, as summarized by Sharma et al..[11] The first strategy is introducing a centrosymmetric-to-noncentrosymmetric structural phase transition in metallic systems such as $LiOsO_3$.[2] The second one is engineering interface-based polar metals in oxide heterostructures.[5] The third one is doping well-established FE insulators and making them into FE metal[12,13]. While all three strategies can potentially create FE materials with large polar or FE distortions, the third strategy is particularly attractive. If one dopes a strong FE insulator to induce metallicity, the system can become a strong FE metal provided that free electrons do not weaken FE distortion. One candidate for applying this strategy is strained $EuTiO_3$, which is a strong FE ferromagnet.[14]

$EuTiO_3$ has attracted wide attention as it exhibits many interesting properties such as multiferroics,[14,15] two-dimensional electron gas,[16,17] superconductivity,[16] ferromagnetism,[18–22] anomalous Hall effect,[23] and giant magnetocaloric effect.[24–26] Bulk $EuTiO_3$ is an antiferromagnetic (AFM) paraelectric (PE) insulator[27–29] with a cubic structure near room temperature. These properties satisfy the criteria to realize FE ferromagnetic (FM) multiferroics proposed by Fennie and Rabe.[15] Indeed, $EuTiO_3$ film becomes a strong FE ferromagnet[14] under strain through spin-lattice coupling. On the other hand, metallicity in $EuTiO_3$ is achieved by doping, which also induces ferromagnetism through the Ruderman-Kittel-Kasuya-Yosida (RKKY) mechanism.[18,20,30] Furthermore, our previous study shows that H-doped strained $EuTiO_3$ is an FE metal in which the polarization can be switched by an external electric field.[31] The rich behaviors of $EuTiO_3$ under strain and doping make it an interesting platform for exploring strong FE metals.

In this paper, we present hybrid density-functional theory (DFT) calculations of the structural, electric, magnetic, and polarization properties of Nb-doped strained $EuTiO_3$ film using the PBE0 functional. A strong FE metal is found in strained $EuTi_{0.875}Nb_{0.125}O_3$ film. The compressive strain field splits the Ti $t_{2g}$ states and shifts up of the $d_{yz}$ and $d_{zx}$ states above the Fermi level. The $d_{xy}$ electrons dominate the Fermi level and do not couple to the polar distortion. The underlying mechanism for the coexistence of metallicity and ferroelectricity in strained $EuTi_{0.875}Nb_{0.125}O_3$ film is revealed in detail. Notably, Nb-doping induces metallicity without weakening ferroelectricity, unlike H-doping, which reduces the ferroelectric displacements. The origin of such difference is discussed and could be useful for guiding future search of strong FE metals.

The hybrid DFT calculations were performed using a tuned PBE0 functional containing 22% Hartree-Fock exchange.[30–32] The projector-augmented wave (PAW) pseudopotentials[33] were used, and the valence states for Eu, Nb, Ti, O, and H include $4f5s5p6s$, $4s4p4d5s$, $3d4s$, $2s2p$, and $1s$ orbitals, respectively. We use a 40-atom supercell to model the films of Nb-doped $EuTiO_3$ ($EuTi_{0.875}Nb_{0.125}O_3$), along with undoped $EuTiO_3$ and H-doped $EuTiO_3$ ($EuTiO_{2.875}H_{0.125}$) for comparison. A 2×2×2 Monkhorst-Pack k-point grid[34] was used to optimize the supercells, and a 5×5×5 Gamma-centered k-point grid was used to calculate the density of states (DOS). The in-plane compressive strain is applied to model the films that are grown along [001] direction on the substrate. As experiments and theoretical calculations have confirmed the G-AFM ground state for undoped $EuTiO_3$ and an FM ground state for bulk $EuTi_{0.875}Nb_{0.125}O_3$[20,30] and

EuTiO$_{2.875}$H$_{0.125}$,[18,31] we only consider the G-AFM and FM magnetic structures for the spin configurations of the strained EuTi$_{0.875}$Nb$_{0.125}$O$_3$, EuTiO$_3$, and EuTiO$_{2.875}$H$_{0.125}$ films. The structural optimization is converged when the total-energy difference between two successive ionic steps is 10$^{-3}$ eV or less. The electronic self-consistent iterations are converged when the total-energy difference between two successive electronic iterations is 10$^{-4}$ eV or less. The plane-wave cutoff energy is 400 eV, and the hybrid DFT calculations were performed with the Vienna Ab initio Simulation Package (VASP).[35] We have used the computational setup described above in our previous works on bulk EuTi$_{0.875}$Nb$_{0.125}$O$_3$,[30] and successfully reproduced the metallic FM ground state observed in experiments.[20]

The strained EuTi$_{0.875}$Nb$_{0.125}$O$_3$, EuTiO$_3$, and EuTiO$_{2.875}$H$_{0.125}$ films have a square-basis lattice parameter $a_f$ which is equal to the lattice parameters of the substrates. The misfit strain is defined as $\eta = (a_f - a_b)/a_b$, where $a_b$ is the in-plane lattice constants of relaxed bulk EuTi$_{0.875}$Nb$_{0.125}$O$_3$, EuTiO$_3$ or EuTiO$_{2.875}$H$_{0.125}$. In the process of structural optimization, the c-axis and the position of all atoms in EuTi$_{0.875}$Nb$_{0.125}$O$_3$, EuTiO$_3$, and EuTiO$_{2.875}$H$_{0.125}$ films are fully relaxed for a given value of $a_f$. Since it is difficult to define the polarization in metallic materials, a physical quantity $P^*$ is defined to characterize the polar distortion in EuTiO$_3$-based FE materials.[31] Here $P^* = (B_l - B_s)/\frac{1}{2}(B_l + B_s)$, where $B_l$ and $B_s$ are the lengths of the long and short Ti-O bonds, respectively. A greater $P^*$ characterizes a larger polar distortion. In addition, a total energy difference $\Delta E_{FE-PE}$ is defined as $E_{FE} - E_{PE}$, where $E_{FE}$ and $E_{PE}$ denote the total energies per f.u. of an FM FE structure and an FM PE structure, respectively. Similarly, a total energy difference $\Delta E_{FM-AFM}$ is defined as $E_{FM} - E_{AFM}$, where $E_{FM}$ and $E_{AFM}$ denote the total energies per f.u. of an FM FE structure and an AFM FE structure, respectively.

First, we present the structural, electronic, magnetic, and polarization properties of Nb-doped EuTiO$_3$ film under compressive strain. Figure 1 shows the values of $\Delta E_{FE-PE}$, $\Delta E_{FM-AFM}$, along with $P^*$ and c/a in the FE phase as a function of misfit strain $\eta$ in EuTi$_{0.875}$Nb$_{0.125}$O$_3$ film. At $\eta = 0$, the total energy of the FE FM state is higher than that of the PE FM state by ~10 meV/f.u.. The system shows PE behavior with Pm-3m symmetry. As compressive strain increases, the behavior of EuTi$_{0.875}$Nb$_{0.125}$O$_3$ film can be divided into two regions: $0 > \eta > -2.1\%$ [marked by region II in Figure 1a] and $-2.1\% \geq \eta \geq -3.0\%$ (region I). EuTi$_{0.875}$Nb$_{0.125}$O$_3$ does not show ferroelectricity with P4/mmm space group at $0 > \eta > -2.1\%$ because the total energy of the polar FE phase is still slightly higher than that of the nonpolar PE phase, as shown in the upper panel of Figure 1a. Also, Figure 1b shows that $P^*$ and c/a increase slowly with the increase of compressive strain in this region. At $\eta = -2.1\%$, the system begins to transform from the P4/mmm phase into the P4mm phase as the total energy of the polar FE phase is lower than the nonpolar PE phase, and $P^*$ and c/a increase suddenly. The energy difference $\Delta E_{FE-PE}$ becomes greater with increasing compressive strain, and $P^*$ and c/a increase rapidly, implying a more stable FE phase in this region. In the whole range of $0 \geq \eta \geq -3.0\%$, EuTi$_{0.875}$Nb$_{0.125}$O$_3$ demonstrates the FM ground state because the FM state has lower total energy than the AFM state, as shown in the lower panel of Figure 1a. At $-2.1 \geq \eta \geq -3.0\%$, the energy difference between the FM and AFM states increases rapidly with the appearance of polarization, indicating spin-lattice coupling in this material system.[14]

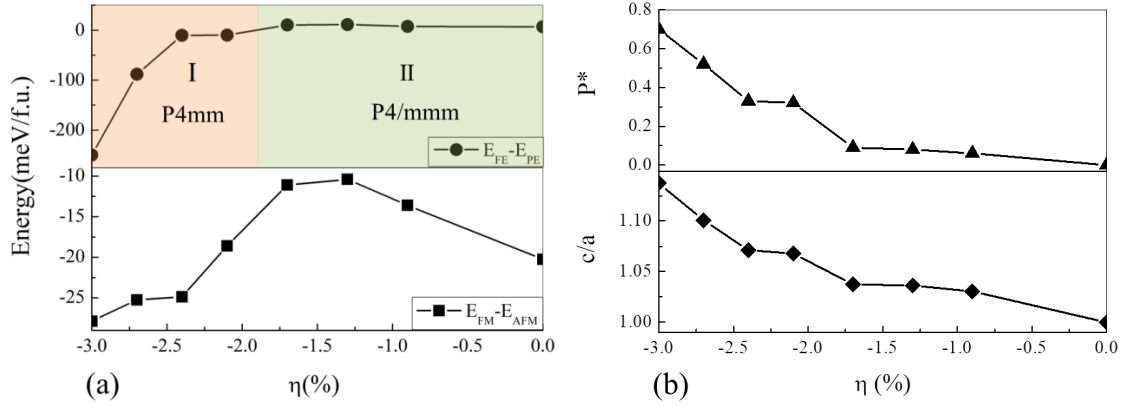

Figure 1. (a) The total energy differences $\Delta E_{FE-PE}$ (upper panel) and $\Delta E_{FM-AFM}$ (lower panel) of strained $EuTi_{0.875}Nb_{0.125}O_3$ at $0 \leq \eta \leq -3\%$ and (b) the polar distortion $P^*$ and c/a, where a and c are the in-plane and out-of-plane lattice constants, respectively.

While the polarization changes significantly with increasing compressive strain, the $EuTi_{0.875}Nb_{0.125}O_3$ film remains metallic at $0 \geq \eta \geq -3.0\%$. Figure 2a and 2b shows the partial DOS of the Ti 3d states in $EuTi_{0.875}Nb_{0.125}O_3$ film at $\eta = 0$ and -3%. The Nb 4d states are not shown here is as it is similar to the Ti 3d states. In both cases, the itinerant Ti 3d states dominate the Fermi surface, which confirms that the metallicity of $EuTi_{0.875}Nb_{0.125}O_3$ film comes from itinerant electrons in Ti 3d and Nb 4d orbitals introduced by Nb dopant. As shown in Figure 1b, a sudden change of $P^*$ and c/a at $\eta = -2.1\%$ implies a well-defined point of structural phase transition to a polar P4mm phase. Therefore, at $\eta \leq -2.1\%$, the strained $EuTi_{0.875}Nb_{0.125}O_3$ film is an FM FE metal. The coexistence of metallicity and ferroelectricity are realized in strained $EuTi_{0.875}Nb_{0.125}O_3$ film, illustrating that combined action of doping and strain engineering is an effective way to obtain new $EuTiO_3$-based FE metal materials.

To understand why metallicity and ferroelectricity coexist in strained $EuTi_{0.875}Nb_{0.125}O_3$, we show the partial DOS (PDOS) of the $EuTi_{0.875}Nb_{0.125}O_3$ film for $\eta = 0$ and -3.0% in Figures 2a and 2b. The strained $EuTi_{0.875}Nb_{0.125}O_3$ remains metallic as the unstrained. However, the characteristics of the states at the Fermi level are changed by the compressive strain. The Ti 3d bands at the Fermi level at $\eta = 0$ are made mainly of $d_{xy}$, $d_{yz}$, $d_{zx}$ states of $t_{2g}$ bands, while the Ti 3d bands at $\eta = -3.0\%$ are made of only the $d_{xy}$ state. The itinerant electron occupies $d_{xy}$, $d_{yz}$, $d_{zx}$ states at $\eta = 0$ because $EuTi_{0.875}Nb_{0.125}O_3$ has a cubic Pm-3m structure that is symmetric in the x-, y-, and z-axes, as shown by the vectors a, b, and c in Figures 2c-2f. As compressive strain increases, the $EuTi_{0.875}Nb_{0.125}O_3$ film transition into the polar P4mm structure, in which the symmetries along the x- and y-axes differ from the one along the z-axis. From Figure 2b, we can see that the strain field splits the $t_{2g}$ bands and induces shifts up the $d_{yz}$ (blue) and $d_{zx}$ (cyan) states above the Fermi level. Meanwhile, the $d_{xy}$ state (red) extends towards lower energies and dominates the Fermi level. As a result, the fraction of itinerant electrons in the $d_{yz}$, $d_{zx}$ orbitals decreases, and the fraction of itinerant electrons in the $d_{xy}$ orbital increases with compressive strain.

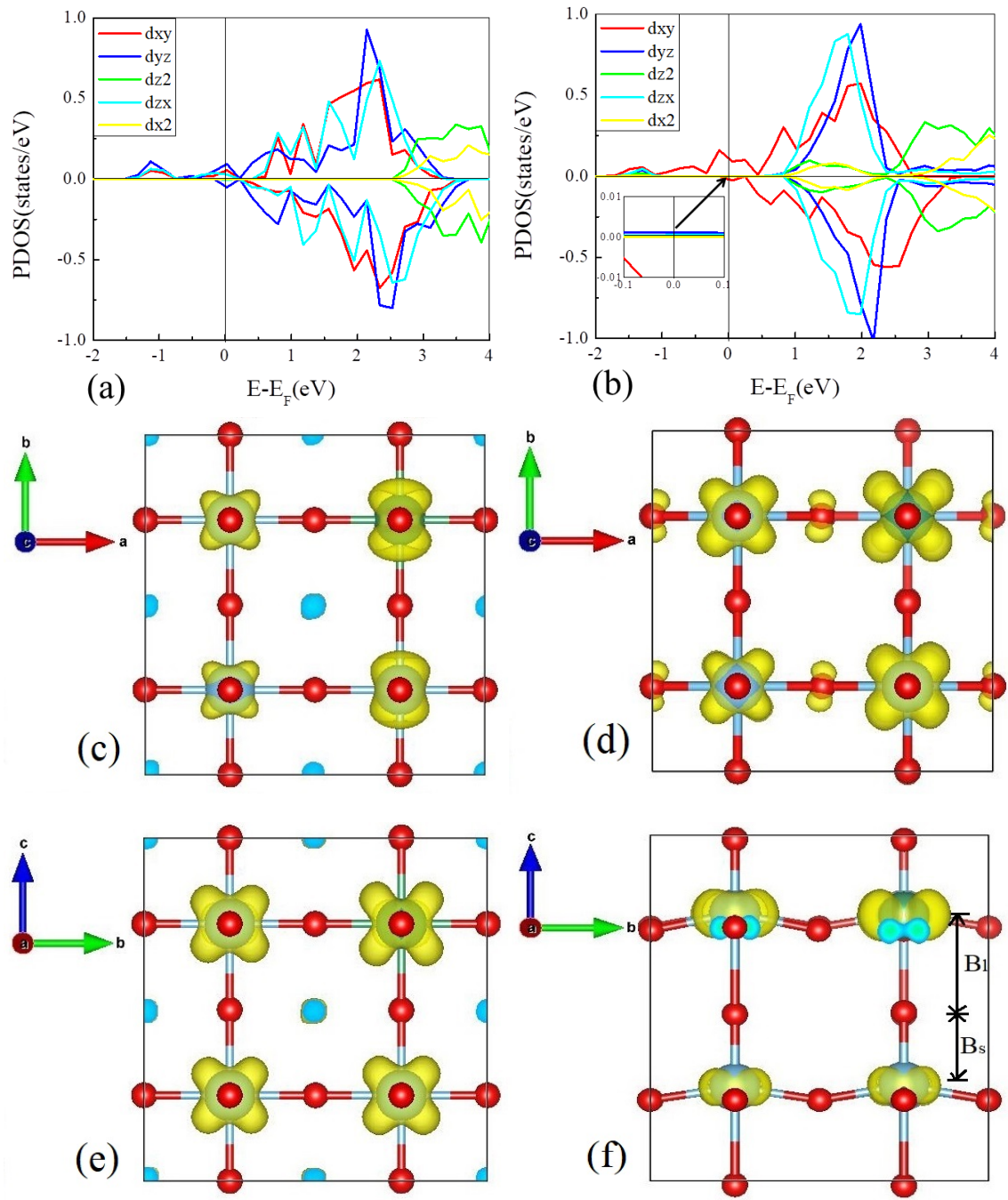

Figure 2. The partial Ti 3d DOS of EuTi$_{0.875}$Nb$_{0.125}$O$_3$ film for (a) $\eta = 0$ and (b) $\eta = -3.0\%$. Top views of band-decomposed charge at the Fermi level for (c) $\eta = 0$ and (d) $\eta = -3.0\%$. Side views of band-decomposed charge at the Fermi level for (e) $\eta = 0$ and (f) $\eta = -3.0\%$. In (f), B$_l$ and B$_s$ are the lengths of the long and short Ti-O bonds, respectively.

It is interesting to investigate why the $d_{xy}$ state extends towards lower energies under the compressive strain. To understand this, in Figures 2c to 2f, we plot the band-decomposed charge density in unstrained and strained $EuTi_{0.875}Nb_{0.125}O_3$ at the Fermi level. At $\eta = -3.0\%$, the polar $EuTi_{0.875}Nb_{0.125}O_3$ film has only $d_{xy}$ orbitals of Ti atoms occupied by itinerant electron (See Figure 2f). Also noticeable is the formation of π bonds between Ti $d_{xy}$ orbitals and O 2p orbitals from the adjacent O atoms in the x-y plane (See Figure 2d). As compressive strain increases, O atoms at the x-y plane become closer to the Ti atoms. As a result, the Ti $d_{xy}$ orbitals and the oxygen p orbitals overlap more, thus enhancing the π bonds. Therefore, the energies of the Ti $d_{xy}$ states in the bonding orbital are lowered (Figure 2b) compared to the unstrained case (Figure 2a). Also, the Ti $d_{xy}$ states become more spatially extended in the x-y plane at the Fermi level (Figure 2d) compared to the unstrained case (Figure 2c), which is consistent with their greater contributions.

The fact that Ti $d_{xy}$ orbitals dominate the Fermi level explains the coexistence of the ferroelectricity and metallicity in strained $EuTi_{0.875}Nb_{0.125}O_3$. According to Anderson and Blount[1], a material can possess these two properties simultaneously if the free electrons are very weakly coupled to the FE distortions. For $EuTi_{0.875}Nb_{0.125}O_3$ films under compressive strain in the x-y plane, the FE distortion mostly manifests as the change of the Ti-O bonds along the z-axis. This distortion is accompanied by secondary effects on the Ti-O bonds in the x-y plane, where the compressive strain causes them to "pucker" along the z-axis. However, the $d_{xy}$ orbitals holding the itinerant electrons are orientated in the (110) and (1-10) directions and do not overlap with any Ti-O bonds either in the z-direction or in the x-y plane. As a result, the itinerant electron in the Ti $d_{xy}$ states at the Fermi level is not coupled to the FE distortion in the polar $EuTi_{0.875}Nb_{0.125}O_3$ film, therefore enabling the coexistence of metallicity and ferroelectricity.

One surprising finding is how strong the ferroelectricity in Nb-doped strained $EuTiO_3$ film is. The strength of ferroelectricity can be evaluated from the polar distortion $P^*$ and the phase stability as reflected in $\Delta E_{FE-PE}$. In Figure 3, we show the $P^*$ and $\Delta E_{FE-PE}$ of strained $EuTi_{0.875}Nb_{0.125}O_3$, $EuTiO_3$, and $EuTiO_{2.875}H_{0.125}$ films for comparison. As compressive strain increases, $P^*$ increases in all three systems, and $\Delta E_{FE-PE}$ decreases. The increase of $P^*$ and decrease of $\Delta E_{FE-PE}$ in the $EuTi_{0.875}Nb_{0.125}O_3$ film are almost the same as that of the $EuTiO_3$ film and are more significant than those of the $EuTiO_{2.875}H_{0.125}$ film. At $\eta = -3.0\%$, the $P^*$ of strained $EuTi_{0.875}Nb_{0.125}O_3$, $EuTiO_3$, and $EuTiO_{2.875}H_{0.125}$ films are 0.322, 0.322 and 0.166, respectively. And $\Delta E_{FE-PE}$ of strained $EuTi_{0.875}Nb_{0.125}O_3$, $EuTiO_3$, and $EuTiO_{2.875}H_{0.125}$ films are 250 meV/f.u., 320 meV/f.u., and 15 meV/f.u., respectively. The polar $EuTi_{0.875}Nb_{0.125}O_3$ film, which has the same structure and symmetry(P4mm) as the strained $EuTiO_3$ film, possesses similar FE distortion and stability for the FE phase. Experimental and theoretical studies[14,15] have shown that the $EuTiO_3$ film under large biaxial compressive strain is a strong FE ferromagnet with spontaneous polarization of 10 μC/cm$^{-2}$. Similarly, the metallic $EuTi_{0.875}Nb_{0.125}O_3$ film under large compressive strain possesses strong ferroelectricity as well. In other words, Nb doping induces metallicity in strained $EuTiO_3$ film while maintaining its strong ferroelectricity.

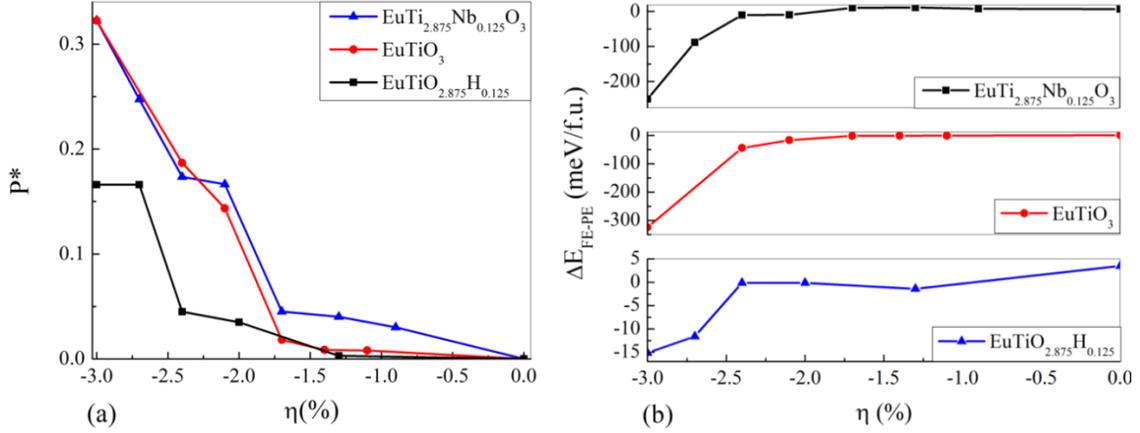

**Figure 3.** (a) The polar distortion P* and (b) the total energy difference of the strained EuTi$_{0.875}$Nb$_{0.125}$O$_3$, EuTiO$_3$, and EuTiO$_{2.875}$H$_{0.125}$ films at $0 \leq \eta \leq -3\%$.

It is interesting that, unlike H-doping, Nb-doping does not weaken the polarization of the strained EuTiO$_3$ film. Previous experimental and theoretical studies[18,20,30,31] show that both Nb- and H-doped bulk EuTiO$_3$ demonstrate metallic ferromagnetism due to the same RKKY interaction mechanism. Why does H-doping induces metallicity and at the same time weakens ferroelectricity in strained EuTiO$_3$ film, but Nb-doping induces metallicity and keeps strong ferroelectricity? To understand the reason for strong ferroelectricity in strained EuTi$_{0.875}$Nb$_{0.125}$O$_3$ film, we show the partial Ti 3d DOS and the band-decomposed charge at the bottom of the conduction band of EuTiO$_3$ film at η = 0 and -3.0% and at the Fermi level of EuTiO$_{2.875}$H$_{0.125}$ at η = -3.0% in Figure 4 and compare them with the EuTi$_{0.875}$Nb$_{0.125}$O$_3$ film in Figure 2. Due to the cubic symmetric structure, the conduction bands of bulk EuTiO$_3$ are made of a combination of Ti $d_{xy}$, $d_{yz}$, $d_{zx}$ states (Figures 4a and 4b). Compressive strain acting on EuTiO$_3$ film induces shifts of the energy of $d_{yz}$ and $d_{zx}$ bands, and the bottom of the conductor bands are made of the Ti $d_{xy}$ state (Figures 4c and 4d). When additional electrons are introduced into the EuTiO$_3$ film by Nb doping, they fill the Ti $d_{xy}$ states and dominate the Fermi level, as shown in Figures 2b and 2d. As we have discussed above, the Ti $d_{xy}$ orbital is not coupled to the polar distortion in compressively strained EuTi$_{0.875}$Nb$_{0.125}$O$_3$. Therefore, Nd-induced free electron at the Fermi level does not weaken ferroelectricity. H-doping also largely fills the Ti $d_{xy}$ bands and induces metallicity in strained EuTiO$_{2.875}$H$_{0.125}$ film. However, there are two important differences between strained EuTiO$_{2.875}$H$_{0.125}$ and EuTi$_{0.875}$Nb$_{0.125}$O$_3$ films. Firstly, the strained EuTiO$_{2.875}$H$_{0.125}$ film has the Pmm2 symmetry, which is different from the P4mm symmetry of strained EuTi$_{0.875}$Nb$_{0.125}$O$_3$ and EuTiO$_3$ films. H-doping breaks the C$_{4v}$ symmetry in strained EuTiO$_3$ and introduces small contributions from Ti $d_{yz}$ and $d_{zx}$ states at the Fermi level in addition to the Ti $d_{xy}$ state (See the insert in Figure 4e). Although the main $d_{xy}$ state is not coupled to the FE distortion, the $d_{yz}$, $d_{zx}$ states are coupled to the strain-induced polarization and thus can weaken the ferroelectricity. In contrast, there are no contributions of the Ti $d_{yz}$ or $d_{zx}$ states at the Fermi level in strained EuTi$_{0.875}$Nb$_{0.125}$O$_3$ (See the insert in Figure 2b). Secondly, unlike the Nb dopant replacing a Ti atom, the H dopant replaces an O atom. After donating the electron to the Ti 3d orbital, the resulting H$^+$ ion generates an attractive Coulombic potential around itself. As a result, the Ti 3d orbitals next to the H$^+$ ion have lower energies compared to those further away and contribute more to the Fermi level, as shown in Figure 4f. In another word, the Ti atoms next to H are closer to the Ti$^{3+}$ (d$^1$) state while the others are closer to Ti$^{4+}$ (d$^0$) state. As discussed in Ref. 32, in strained EuTiO$_{2.875}$H$_{0.125}$ film, the Ti$^{3+}$ (d$^1$) ions are responsible for metallicity, and Ti$^{4+}$ (d$^0$) are responsible for ferroelectricity. When H-doping introduces metallicity, it converts some Ti$^{4+}$ (d$^0$) ions to Ti$^{3+}$ (d$^1$) ions, which inevitably weakens the ferroelectricity as the number of Ti$^{4+}$ (d$^0$) ions is reduced. The overall results suggest that doping at the Ti sites will be more beneficial than doping at the O site for obtaining strong FE metals in EuTiO$_3$ and similar material systems.

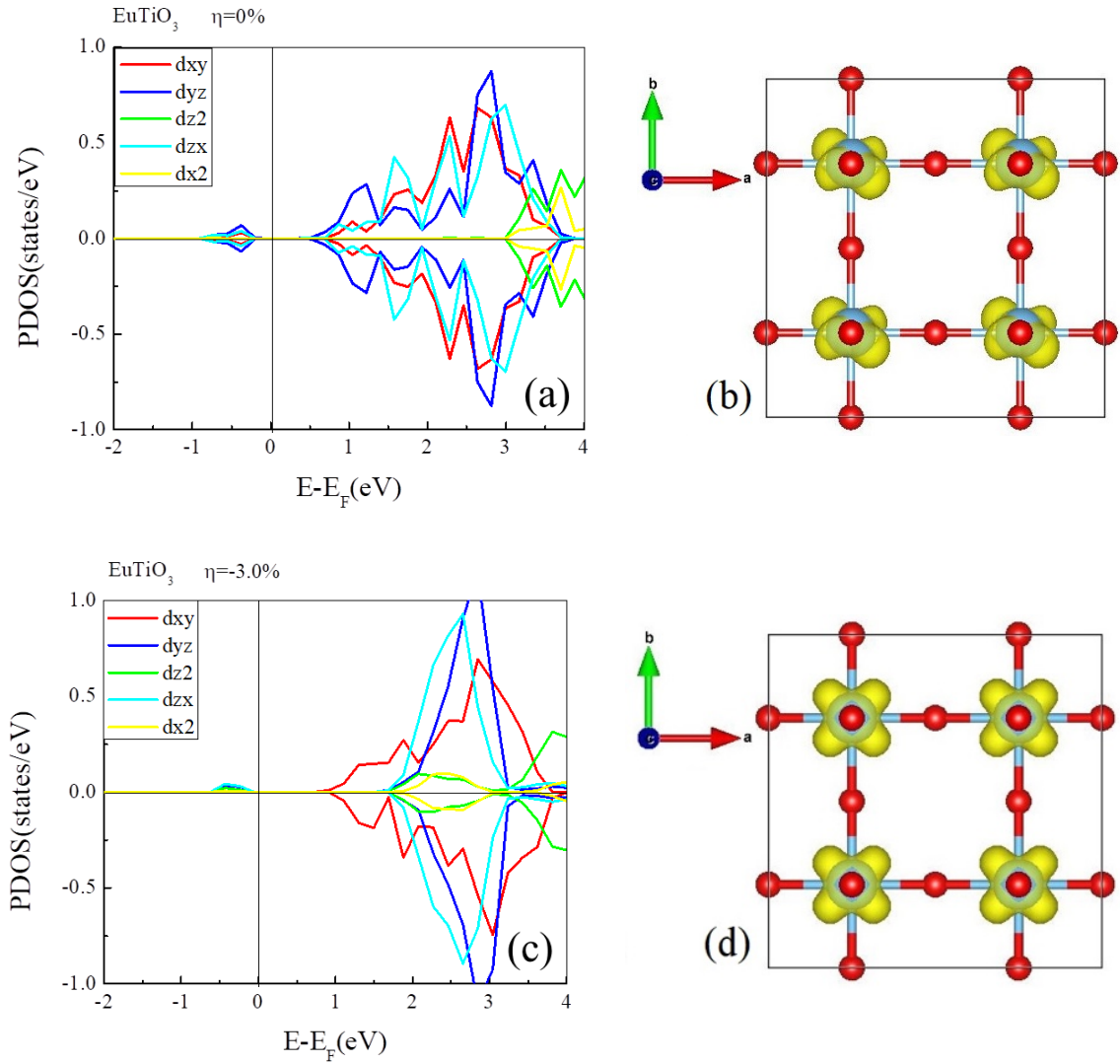

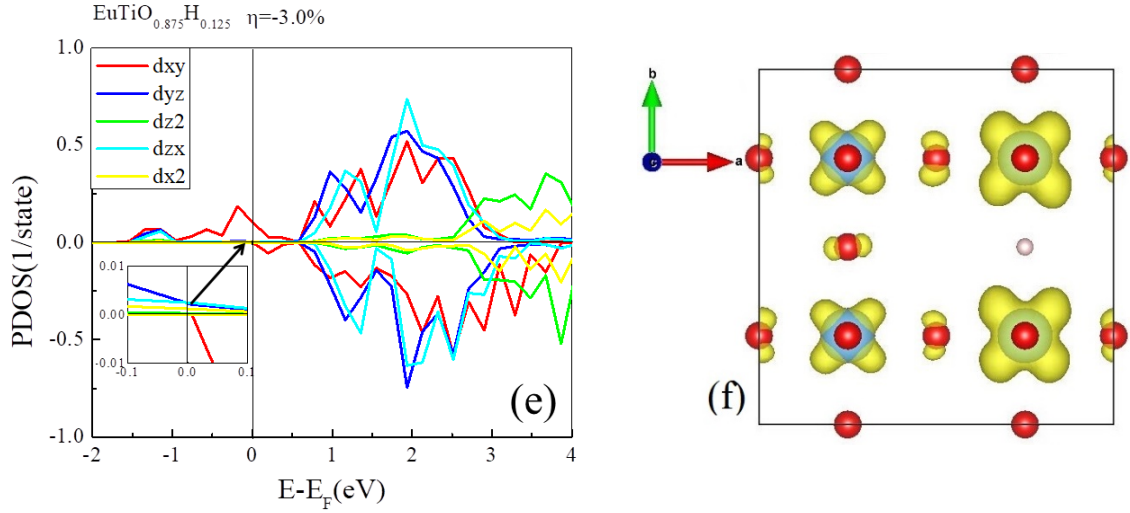

Figure 4. The partial Ti 3d DOS (a) and the band-decomposed charge at the bottom of the conduction band (b) for EuTiO$_3$ film at η = 0. The partial Ti 3d DOS (c) and the band-decomposed charge at the Fermi level (d) for EuTiO$_3$ film at η = -3.0%. The partial Ti 3d DOS (e) and the band-decomposed charge at the Fermi level (f) for EuTiO$_{2.875}$H$_{0.125}$ at η = -3.0%.

In conclusion, we investigated the structural, electronic, and polarization properties of strained EuTi$_{0.875}$Nb$_{0.125}$O$_3$ by hybrid DFT calculations. The strained EuTi$_{0.875}$Nb$_{0.125}$O$_3$ film is a PE FM metal at 0 ≥ η > -2.1% and becomes a strong FE FM metal at η ≤ −2.1%. The compressive strain splits the t$_{2g}$ bands, with the d$_{yz}$, d$_{zx}$ states shifting up, and the d$_{xy}$ states shifting down and dominating the Fermi level. Due to their geometry, the d$_{xy}$ orbitals are not coupled to the FE distortion, which explains the coexistence of metallicity and strong ferroelectricity. We also discussed why Nb doping does not weaken FE distortions in strained EuTiO$_3$ unlike H-doping. The results demonstrate that combining strain engineering and doping, especially at the Ti sites, is effective in achieving EuTiO$_3$-based strong FE metals and may also be useful for other material systems.

Acknowledgement

S. Xu and Y. Gu are thankful for the financial support from the Doctoral Research Project of JUST (Nos. JKD120114001). We are grateful to Jiangsu University of Science and Technology (S. Xu and Y. Gu) and the high-performance computing facility at the University of Memphis (X. Shen) for the award of CPU hours.